\documentclass[prl,twocolumn,showpacs,floatfix,superscriptaddress]{revtex4}
\input{psfig.sty}
\usepackage{graphicx}
\def\mpl{\ifmmode \overline M_{Pl}\else $\overline M_{Pl}$\fi}

\newcommand{\pom}{I\!\! P}

\begin{document}
\bibliographystyle{revtex}
\input{psfig.sty}
\title{Inclusive Double Pomeron Exchange at the Fermilab Tevatron $\bar pp$ Collider\\
\vglue 0.15in
The CDF Collaboration\\
\vglue 0.15in
(Submitted to Physical Review Letters)}

\font\eightit=cmti8
\def\r#1{\ignorespaces $^{#1}$}
\hfilneg
\author{
\noindent
D.~Acosta,\r {14} T.~Affolder,\r 7 H.~Akimoto,\r {51}
M.G.~Albrow,\r {13} D.~Ambrose,\r {37}   
D.~Amidei,\r {28} K.~Anikeev,\r {27} J.~Antos,\r 1 
G.~Apollinari,\r {13} T.~Arisawa,\r {51}
% ...
}
\affiliation{\r 1  {\eightit Institute of Physics, Academia Sinica, Taipei, Taiwan 11529, 
Republic of China} \\
\r 2  {\eightit Argonne National Laboratory, Argonne, Illinois 60439} \\
\r 3  {\eightit Istituto Nazionale di Fisica Nucleare, University of Bologna,
I-40127 Bologna, Italy} \\
\r 4  {\eightit Brandeis University, Waltham, Massachusetts 02254} \\
\r 5  {\eightit University of California at Davis, Davis, California  95616} \\
\r 6  {\eightit University of California at Los Angeles, Los 
Angeles, California  90024} \\ 
\r 7  {\eightit University of California at Santa Barbara, Santa Barbara, California 
93106} \\ 
\r 8 {\eightit Instituto de Fisica de Cantabria, CSIC-University of Cantabria, 
39005 Santander, Spain} \\
\r 9  {\eightit Carnegie Mellon University, Pittsburgh, Pennsylvania  15213} \\
\r {10} {\eightit Enrico Fermi Institute, University of Chicago, Chicago, 
Illinois 60637} \\
\r {11}  {\eightit Joint Institute for Nuclear Research, RU-141980 Dubna, Russia}
\\
\r {12} {\eightit Duke University, Durham, North Carolina  27708} \\
\r {13} {\eightit Fermi National Accelerator Laboratory, Batavia, Illinois 
60510} \\
\r {14} {\eightit University of Florida, Gainesville, Florida  32611} \\
\r {15} {\eightit Laboratori Nazionali di Frascati, Istituto Nazionale di Fisica
               Nucleare, I-00044 Frascati, Italy} \\
\r {16} {\eightit University of Geneva, CH-1211 Geneva 4, Switzerland} \\
\r {17} {\eightit Glasgow University, Glasgow G12 8QQ, United Kingdom}\\
\r {18} {\eightit Harvard University, Cambridge, Massachusetts 02138} \\
\r {19} {\eightit Hiroshima University, Higashi-Hiroshima 724, Japan} \\
\r {20} {\eightit University of Illinois, Urbana, Illinois 61801} \\
\r {21} {\eightit The Johns Hopkins University, Baltimore, Maryland 21218} \\
\r {22} {\eightit Institut f\"{u}r Experimentelle Kernphysik, 
Universit\"{a}t Karlsruhe, 76128 Karlsruhe, Germany} \\
\r {23} {\eightit High Energy Accelerator Research Organization (KEK), Tsukuba, 
Ibaraki 305, Japan} \\
\r {24} {\eightit Center for High Energy Physics: Kyungpook National
University, Taegu 702-701; Seoul National University, Seoul 151-742; and
SungKyunKwan University, Suwon 440-746; Korea} \\
\r {25} {\eightit Ernest Orlando Lawrence Berkeley National Laboratory, 
Berkeley, California 94720} \\
\r {26} {\eightit University College London, London WC1E 6BT, United Kingdom} \\
\r {27} {\eightit Massachusetts Institute of Technology, Cambridge,
Massachusetts  02139} \\   
\r {28} {\eightit University of Michigan, Ann Arbor, Michigan 48109} \\
\r {29} {\eightit Michigan State University, East Lansing, Michigan  48824} \\
\r {30} {\eightit Institution for Theoretical and Experimental Physics, ITEP,
Moscow 117259, Russia} \\
\r {31} {\eightit University of New Mexico, Albuquerque, New Mexico 87131} \\
\r {32} {\eightit Northwestern University, Evanston, Illinois  60208} \\
\r {33} {\eightit The Ohio State University, Columbus, Ohio  43210} \\
\r {34} {\eightit Osaka City University, Osaka 588, Japan} \\
\r {35} {\eightit University of Oxford, Oxford OX1 3RH, United Kingdom} \\
\r {36} {\eightit Universita di Padova, Istituto Nazionale di Fisica 
          Nucleare, Sezione di Padova, I-35131 Padova, Italy} \\
\r {37} {\eightit University of Pennsylvania, Philadelphia, 
        Pennsylvania 19104} \\   
\r {38} {\eightit Istituto Nazionale di Fisica Nucleare, University and Scuola
               Normale Superiore of Pisa, I-56100 Pisa, Italy} \\
\r {39} {\eightit University of Pittsburgh, Pittsburgh, Pennsylvania 15260} \\
\r {40} {\eightit Purdue University, West Lafayette, Indiana 47907} \\
\r {41} {\eightit University of Rochester, Rochester, New York 14627} \\
\r {42} {\eightit Rockefeller University, New York, New York 10021} \\
\r {43} {\eightit Instituto Nazionale de Fisica Nucleare, Sezione di Roma,
University di Roma I, ``La Sapienza," I-00185 Roma, Italy}\\
\r {44} {\eightit Rutgers University, Piscataway, New Jersey 08855} \\
\r {45} {\eightit Texas A\&M University, College Station, Texas 77843} \\
\r {46} {\eightit Texas Tech University, Lubbock, Texas 79409} \\
\r {47} {\eightit Institute of Particle Physics, University of Toronto, Toronto
M5S 1A7, Canada} \\
\r {48} {\eightit Istituto Nazionale di Fisica Nucleare, University of Trieste/\
Udine, Italy} \\
\r {49} {\eightit University of Tsukuba, Tsukuba, Ibaraki 305, Japan} \\
\r {50} {\eightit Tufts University, Medford, Massachusetts 02155} \\
\r {51} {\eightit Waseda University, Tokyo 169, Japan} \\
\r {52} {\eightit University of Wisconsin, Madison, Wisconsin 53706} \\
\r {53} {\eightit Yale University, New Haven, Connecticut 06520} \\
}
\collaboration{CDF Collaboration}
\noaffiliation
\begin{abstract}
We report a study of 
$\bar p+p\rightarrow \bar p+X+Y$ at $\sqrt s=1800$ GeV,
where $Y$ is a proton or system of mass-squared $M_Y^2\lesssim 8$ GeV$^2$.
In a sample of events with a leading $\bar p$ of fractional momentum loss 
$0.035<\xi_{\bar p}<0.095$ and 4-momentum transfer squared 
$|t_{\bar p}|<1.0$ GeV$^2$, the proton fractional momentum loss $\xi_p^X$ 
to the system $X$ was evaluated from the momenta of 
the particles comprising $X$.
In the region $\xi_p^X<0.02$, the $\xi_p^X$ distribution behaves as 
$\sim 1/(\xi_p^X)^{1.1}$, as expected for double Pomeron exchange. 
The fraction of events with $\xi_p^X<0.02$ is found to 
be $0.194\pm 0.001\,{\rm (stat)}\pm 0.012\,{\rm (syst)}$. 
\end{abstract}

\maketitle
The success of perturbative Quantum Chromo-Dynamics (QCD)  
in describing strong interactions at high transverse momentum transfers 
rests on the factorization theorem, which 
allows hadronic cross sections to be expressed in terms of parton-level 
cross sections convoluted with uniquely defined hadron parton densities.
It is therefore not surprising that the recently reported
breakdown of factorization in diffractive dijet 
production~\cite{CDF_JJ_RP_1800}, 
a process containing both a hard scattering and the characteristic 
rapidity gap signature of diffraction, has attracted 
considerable theoretical attention. 
Rapidity gaps,  defined as regions of pseudorapidity~\cite{pseudo}
devoid of particles, are  presumed to be formed in diffractive events by 
the exchange of Pomerons ($\pom$), which in QCD correspond to entities of gluons 
and/or quarks with the quantum numbers of the vacuum~\cite{Regge} 
(see Fig.~1). 
The breakdown of factorization in diffraction is generally attributed to 
a suppression of the cross section
resulting from  additional partonic interactions within a diffractive event 
that spoil the rapidity gap signature~\cite{GLM,KKMR}.  
In processes with two rapidity gaps, as in that with two forward gaps 
traditionally referred to as  
double Pomeron exchange (DPE), shown in Fig.~1b, 
one might then expect that either both gaps survive or are simultaneously 
spoiled, leading to a largely non-suppressed ratio of two-gap 
to one-gap rates~\cite{multigap}.
Such a scenario could explain our finding that the ratio
of the rates of DPE to  single diffractive (SD) dijet production 
is about five times larger than that of SD
to non-diffractive (ND) dijet production~\cite{CDF_DPE}.
\begin{figure}
\centerline{\psfig{figure=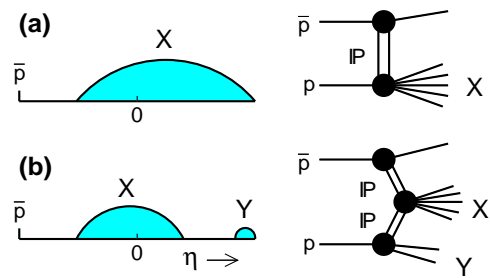,width=2.5in}}
\caption{Diagrams and event topologies for (a) single diffraction,
$\bar p+p\rightarrow \bar p+X$, 
and (b) double Pomeron ($\pom$) exchange, $\bar p+p\rightarrow \bar p+X+Y$; 
the shaded areas represent pseudorapidity regions of particle production.}
\end{figure}

Since rapidity gap formation is a non-perturbative phenomenon, 
soft (low transverse momentum) 
diffractive cross sections would be expected to 
exhibit a similar behavior. Indeed, the SD
$\bar pp$ cross section has been 
found to be suppressed at high energies by a factor of $\sim 10$ 
relative to extrapolations from lower energy data based
on Regge theory and factorization~\cite{CDF_PRD,R,GM}. 
In this Letter, we present a measurement of the ratio of the inclusive DPE 
to SD cross sections in $\bar pp$ collisions at $\sqrt s=1800$ GeV
and compare our results with previous measurements~\cite{UA8} and with 
predictions from Regge theory and various
theoretical models proposed to account for the breakdown of 
Regge factorization in SD. 
Our measurement
severely constrains the available models,
paving the way towards a more comprehensive understanding of the physics of 
rapidity gaps.

The components of the Collider Detector at Fermilab (CDF) 
relevant to this study 
are the Roman Pot Spectrometer (RPS)~\cite{CDF_JJ_RP_1800}, used to detect 
leading antiprotons, and the calorimeters and beam-beam counters 
(BBC)~\cite{CDF}, used to detect the particles from proton dissociation. 
The RPS measures the fractional momentum loss $\xi_{\bar p}$ 
and 4-momentum transfer squared $t_{\bar p}$ of the antiproton 
with resolutions 
$\delta\xi_{\bar p}=\pm 1.0\times 10^{-3}$ and 
$\delta t_{\bar p}=\pm 0.07$ GeV$^2$, respectively~\cite{CDF_JJ_RP_1800}. 
The calorimeters have projective tower geometry and cover the
regions $|\eta|<1.1$ (central), $1.1<|\eta|<2.4$ (plug), and $2.2<|\eta|<4.2$
(forward). The $\Delta \eta \times \Delta \phi$ tower dimensions are
approximately $0.1\times 15^{\circ}$ for the central 
and $0.1\times 5^{\circ}$ for the plug and forward calorimeters. 
The BBC consist of two arrays of eight vertical and eight 
horizontal scintillation counters perpendicular to the beam line 
at $z=\pm 6$~m, BBC$_{\bar p}$ 
and BBC$_p$, covering approximately the region $3.2<|\eta|<5.9$
in four $\eta$-segments of width $\Delta\eta\approx 0.7$.

The present study is based on our $\sqrt s=$1800 
GeV inclusive SD data sample~\cite{CDF_JJ_RP_1800}.
The events were collected in the 1995-96 Tevatron Run 1C 
by triggering on an antiproton detected in the RPS. 
Offline cuts were applied requiring a reconstructed track in the RPS,
no more than one reconstructed vertex in the CDF detector
within a distance $|z_{\rm vtx}|<60$ cm from the nominal beam-beam interaction 
point along the beam direction, and a BBC$_{\bar p}$ multiplicity of $\leq 6$. 
These cuts remove overlap events due to 
multiple interactions in the same beam-beam crossing, comprising
4\% of the inclusive SD data sample as estimated by the 
instantaneous luminosity. 

Experimentally, we study the DPE process $\bar p+p\rightarrow \bar p+X+Y$, 
where $Y$ is either a leading proton or a low-mass proton dissociation system
which escapes undetected through the beam-pipe on the proton side; the 
mass-squared of the system $Y$ is estimated to be $M_Y^2\lesssim 8$ GeV$^2$.
The procedure we follow to
identify and measure the DPE signal in these data is to select 
an event sample with ($\xi_{\bar p},\,t_{\bar p}$) within a certain 
region and measure the fractional longitudinal momentum 
$\xi_p^X$ of the proton transferred 
to the system $X$ using the equation 
\begin{equation}
\xi_p^X=\frac{1}{\sqrt{s}}\sum_{i=1}^nE_T^ie^{\eta^i},
\label{xi}
\end{equation}
where $E_T^i$ and $\eta^i$ are the transverse energy and pseudorapidity 
of a particle~\cite{pseudo} and 
the sum is carried out over all particles excluding the proton or the 
particles associated with 
the system $Y$. 
DPE events are expected to appear in the 
low $\xi_p^X$ region, in contrast to SD events for which $\xi_p^X\approx 1$. 
In practice, not all particles of the system $X$ are included 
in evaluating Eq.~(1) because (a) CDF does not provide full coverage 
and (b) particles depositing energy in the calorimeters below the 
energy thresholds used to reject noise are excluded. 
This issue is addressed by applying appropriate correction factors and by
calibrating formula (1) on
the antiproton side by directly comparing the value of $\xi_{\bar p}$ obtained 
by this method with that measured by the RPS, $\xi_{\bar p}^{RPS}$, 
as discussed below. 

To evaluate $\xi_p^X$ we use calorimeter towers and BBC hits.
The tower energy thresholds used, chosen to lie
comfortably above noise level, are $E_T=0.3$ GeV for the central, 
$E_T=0.2$ GeV for the plug, and $E=1.5$ GeV for the forward calorimeters; 
at the calorimeter interface near $|\eta|\sim 2.4$ a 
threshold of $E_T=0.275$ GeV was used. 
These values are based on test-beam calibrations of the 
calorimeters~\cite{CDF} 
and must be multiplied by an $\eta$-dependent factor $f_{E_T}$ 
(of average value $\langle f_{E_T}\rangle=1.6$)
to obtain the true $E_T$ at low energies~\cite{Suren}.
To account for particles below tower threshold, the calorimeter contribution 
to $\xi_p^X$ is multiplied by $f_{\rm thr}=1.54$. This factor is obtained
from a Monte Carlo (MC) simulation
in which the same tower thresholds are used as in the data after
dividing the generated particle energy by $f_{E_T}$.
The Monte Carlo simulation is based on the single diffractive 
generator described in~\cite{CDF_PRD} and references therein, adapted 
to double Pomeron exchange.
For each BBC hit we use $\eta$ and $E_T$ values randomly chosen 
from a flat $\eta$ distribution over the hit BBC $\eta$-segment 
and from the shape of the $E_T$ distribution expected from the MC simulation,
respectively. The BBC contribution to $\xi_p^X$ 
is then weighted by a factor of 3/2 to account 
for neutral particles, which are undetected by the BBC, and by an additional 
factor of 3/4 to account for the overlap regions among the four 
scintillation counters of each BBC segment. Hits in the outer 
$\eta$-segments, $3.2<|\eta|<3.9$, which overlap with the forward calorimeters,
are ignored. The BBC contribution to $\xi_p^X$ is less than 10\% in the 
region of $-4<\log \xi_p^X<-2$ and increases to $60\%$ at $\log \xi_p^X=-5$
and $\log \xi_p^X=-1$.

	The method of measuring $\xi$ using Eq.~(1) is calibrated on 
the antiproton side by evaluating
$\xi_{\bar p}^X\equiv \frac{1}{\sqrt{s}}\sum_{i=1}^nE_T^ie^{-\eta^i}$ 
(excluding the antiproton from the sum) 
and comparing its value  with that measured by the RPS.
The data are divided into bins of $\Delta\xi_{\bar p}^{RPS}=0.01$, and 
the $\xi_{\bar p}^X$ values obtained for each bin are fitted with a 
Landau distribution.
Figure~2a shows, as an example,  
the data and fit for $0.05<\xi_{\bar p}^{RPS}<0.06$.
The ratio of width to peak position is $\approx 0.6$ over 
the entire $\xi_{\bar p}$ region of our data sample. 
The enhancement in the small $\xi_{\bar p}^X$ region is 
caused by a downward shift in $\xi_{\bar p}^X$ in low multiplicity events 
due to ``loss'' of particles with energy under tower threshold.
Within the region $0.01<\xi_{\bar p}^{RPS}<0.1$, an 
approximately linear relationship 
is observed between the median value of 
${\xi}_{\bar p}^X$  and $\xi_{\bar p}^{RPS}$. A fit with 
$\bar{\xi}_{\bar p}^X \approx C \xi_{\bar p}^{RPS}$ yields 
$C=0.95$, in close agreement with the expected value $C=1$.
A fit in which $C\equiv 1$ and $\langle f_{E_T}\rangle$ is varied with
$f_{\rm corr}\equiv \langle f_{E_T}\rangle\times f_{\rm thr}$
treated as a free parameter yields $f_{\rm corr}=2.7$. In Fig.~2b
an error of $\pm 5\%$ is used in all data points 
to yield $\chi^2/{\rm d.o.f.}=1$ for this fit. In extracting results, 
we use $f_{\rm corr}=2.7$ and assign 
a conservative $\pm 10\%$ error to 
$C$ (twice the error obtained from the fit) to account for other 
possible systematic uncertainties.

\begin{figure}[htb]
\centerline{\psfig{figure=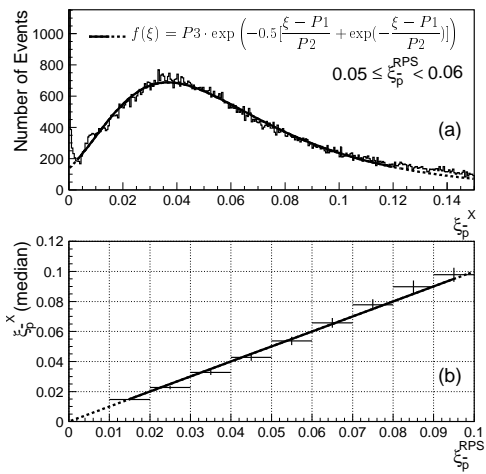,width=2.5in}}
\caption{(a) Distribution of antiproton fractional momentum loss 
$\xi_{\bar p}^X$ measured from calorimeter and beam-beam counter information
for events in which the $\xi_{\bar p}^{RPS}$ value measured by the Roman Pot 
Spectrometer is within $0.05<\xi_{\bar p}^{RPS}<0.06$; the solid line is 
a Landau fit. 
(b) Median values $\bar{\xi}_{\bar p}^X$ obtained from Landau fits to data in 
different $\xi_{\bar p}^{RPS}$ bins plotted versus $\xi_{\bar p}^{RPS}$;
a linear relationship is observed.}
\end{figure}

The DPE signal is evaluated for events 
with antiproton $\xi_{\bar p}$ and $t_{\bar p}$  
within $0.035<\xi_{\bar p}<0.095$ and $|t_{\bar p}|<1.0$ GeV$^2$, where 
the RPS acceptance is larger than $\approx 30\%$~\cite{CDF_JJ_RP_1800}.  
The total number of inclusive SD events in this region is 568K.
The calibrated $\xi_p^X$ distribution is compared in Fig.~3 with 
a two-component MC simulation that includes SD and DPE.
The shape of the input $\xi_p$ distribution in the MC simulation 
for DPE is based on a
triple-Pomeron term on the proton side using a Pomeron intercept 
$\alpha_{\pom}(0)=1+\epsilon$ with $\epsilon=0.104$, as determined 
from a global fit to $p(\bar p)$ total cross section data~\cite{CMG}. 
DPE events were generated for $\xi_p<0.1$.
The DPE and SD MC generated events are independently normalized to the 
data points in the regions $4\times 10^{-5}<\xi_p^X<10^{-2}$ and 
$0.02<\xi_p^X<1$, respectively.
The SD events appear as a broad peak around $\xi_p^X=1$, which falls 
exponentially as $\xi_p^X$ decreases. The DPE events appear as a flattening
of the distribution on the low $\xi_p^X$ side and represent the dominant 
contribution for $\xi_p^X<0.02$.  
The wavy shape of the data distribution in the DPE region is due to 
the $\eta$-dependent calorimeter tower 
energy thresholds used and is reproduced by the MC simulation.  
At low $\xi_p^X$ both data and MC simulation extend down to 
and below the kinematic limit of 
$\xi_{p,\rm min}={M_0^2}/(s\xi_{\bar p,\rm min})\approx 10^{-5}$, 
where $M_0$ is the lowest mass for DPE excitation after threshold 
turn-on effects set in, taken to be 1~GeV.
The events below the kinematic limit are due to the downward 
fluctuations of $\xi_p^X$ in low multiplicity events mentioned above. 
The agreement between data and MC simulation in the region of 
$\xi_p^X<0.02$ shows that Regge factorization 
is successful in describing the shape of the $\xi$ distribution 
in DPE using the Pomeron intercept determined in~\cite{CMG}.
\begin{figure}[htp]
\centerline{\psfig{figure=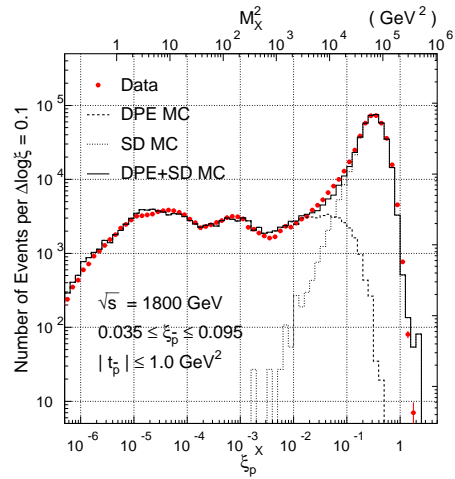,width=2.5in}}
\caption{Distribution of proton fractional momentum loss
$\xi_{p}^X$, measured from calorimeter and beam-beam counter information,
for events with a leading antiproton of $0.035<\xi_{\bar p}^{RPS}<0.095$ 
and $|t_{\bar p}|<1.0$ GeV$^2$; the curves are from a Monte Carlo 
simulation of SD (dotted), DPE (dashed) and total (solid) contributions 
normalized to the data points; the DPE events were generated for 
$\xi_p<0.1$.}
\end{figure}     
The ratio of the number of events within $\xi_p^X<0.02$ to the total 
number of events is $0.202\pm 0.001\,{\rm (stat)}$.
After correcting for smearing effects caused by the $\xi_p^X$ resolution, 
the ratio becomes $R^{DPE}_{SD}=0.194\pm 0.001\,{\rm (stat)}\pm 0.012\,{\rm (syst)}$, where the systematic error is from the uncertainties due to   
$\xi_p^X$ calibration ($\pm 0.003$), $\xi_p^X$ smearing ($\pm 0.008$), and 
low $\xi_p^X$ enhancement ($\pm 0.008$, see Fig.~2a) added in quadrature. 

Neglecting Reggeon contributions, the DPE/SD ratio is given in 
Regge theory by~\cite{multigap}
\begin{equation}
R^{DPE}_{SD}|_{\xi_{\bar p}}=\int_{t_p=-\infty}^{0}
\int_{\xi_{p,\rm min}}^{0.02}\frac{\kappa\beta^2(t_p)dt_pd\xi_p}
{16\pi\xi_p^{\alpha(0)+2\alpha't_p}}
\label{reggeratio}
\end{equation}
where $\kappa$ is the ratio of the triple-Pomeron coupling $g(t_p)$ 
to the Pomeron-proton coupling $\beta(t_p)$ and $\alpha(t)=
\alpha(0)+\alpha't$ is the Pomeron trajectory. Using $\kappa=0.170\pm 0.017$, 
$\beta(t_p)=\beta(0)e^{4.6t_p}$, $\beta^2(0)/16\pi=0.86$ GeV$^{-2}$
and $\alpha(t)=1.104+0.25t$~\cite{GM} yields 
$R^{DPE}_{SD}({\rm Regge})=0.36\pm 0.04$. This prediction 
is larger than the measured values by a factor 
of $1.9\pm 0.2$. However, this discrepancy from the factorization 
expectation of unity is small compared to 
the ${\cal{O}}(10)$ discrepancy  observed in SD~\cite{R}.
Thus, this result confirms the conjecture made in the introduction 
that the formation of a rapidity gap within the rapidity space covered by 
the diffraction dissociation products in events with a leading (anti)proton
would be largely non-suppressed. A similar conclusion has been reached by 
the UA8 Collaboration from a study of DPE production 
in $\bar p p$ collisions at $\sqrt s=$630 GeV at the CERN $S\bar ppS$
collider~\cite{UA8}. 
Changes in the predicted two-gap to one-gap ratio due to 
contamination of the DPE signal with proton fragmentation events 
are estimated to be $\sim 15\%$ and therefore are not expected to 
alter this conclusion.

Phenomenological models proposed to account for the
breakdown of Regge factorization in SD may be divided into two 
broadly defined classes: (a) those attributing the violation either to 
``damping" of the cross section at small $\xi$~\cite{ES98}
or to a decrease of the Pomeron intercept at low $\xi$~\cite{tan}
or at high energies~\cite{ES00}, and (b) those in which the overall 
normalization decreases with increasing energy but the shape of the 
$\xi$ distribution remains practically~\cite{GLM,KKMR,GLM2} or 
entirely~\cite{multigap,R} unchanged. The models of class (a) 
predict a $\xi_p^X$ distribution different from that expected from SD and 
are disfavored by the shape of the distribution presented in Fig.~3, 
which behaves as $1/\xi^{\alpha(0)}$ down to the kinematic limit of 
$\xi_{p,\rm min}\approx 10^{-5}$. Of the class (b) models, three  have reported
predictions for both SD and DPE: the eikonal model~\cite{GLM2}, 
the Pomeron flux renormalization model~\cite{R}, and the gap probability 
renormalization model~\cite{multigap}. The eikonal model, 
in which ``screening corrections'' to the Regge 
amplitude are calculated using an eikonal approach, yields 
suppression factors of 0.369 and 0.309 for SD and DPE, respectively; 
although the DPE/SD ratio is relatively non-suppressed, in close agreement 
with our result, the suppression for SD is 
underestimated by a factor of $\sim 3$. 
The Pomeron flux renormalization model, in which the Regge theory  
Pomeron flux factor is renormalized to unity for Pomerons 
emitted by the $\bar p$ in SD or DPE and independently by the $p$ in DPE, 
yields the correct suppression factor for SD, but predicts a 
DPE/SD ratio smaller than the measured value by a 
factor of $4.7\pm 0.6$~\cite{R}. 
Finally, in the gap probability renormalization model, 
in which the SD and DPE cross sections are expressed in terms of the 
variables $M^2_X$ and $\Delta\eta=\Delta\eta_{\bar p}+\Delta\eta_p$,
where $\Delta\eta_i=-\ln\xi_i$, the predicted DPE/SD ratio is 
$0.21\pm 0.02$~\cite{multigap}, in good agreement with our
measured value of $0.194\pm 0.001\pm 0.012$. These predictions do not include 
possible effects from Reggeon exchange or contributions from 
proton fragmentation.  

In summary, we have studied the double Pomeron exchange (DPE) process 
$\bar p+p\rightarrow \bar p+X+Y$, where $Y$ is a leading proton or 
a proton-dissociation system of mass-squared $M_Y^2\lesssim 8$ GeV$^2$,
by measuring the fractional longitudinal momentum loss of the proton 
to the system $X$, $\xi_p^X$, 
in events with a leading antiproton of 
$0.035<\xi_{\bar p}<0.095$ and $|t_{\bar p}|<1.0$ GeV$^2$ 
produced in $\bar pp$ 
collisions at $\sqrt s=$1800 GeV. Events in the region  
$\xi_p^X<0.02$ follow a distribution of the form 
$\sim 1/\xi_p^{\alpha(0)}$, where $\alpha(0)$ is the 
Pomeron intercept, and are attributed to DPE production. 
The ratio of the number of DPE 
events in this region to the total number of SD events is found to be 
$0.194\pm 0.001\pm 0.012$. This value is lower than the  
prediction based on Regge factorization by a factor of $1.9\pm0.2$,
which is relatively small compared to the suppression factor 
of ${\cal{O}}(10)$ observed in SD~\cite{R},
indicating that the formation of a second rapidity gap in a SD event 
is relatively non-suppressed. Among models proposed to explain the suppression 
of the SD cross section at high energies, our results favor those 
in which the Regge based shapes of the SD and DPE distributions 
remain unchanged and only the overall normalization is 
suppressed~\cite{GLM,KKMR,multigap,R,GLM2}. 

We thank the Fermilab staff and the technical staffs of the
participating institutions for their vital contributions.  This work was
supported by the U.S. Department of Energy and National Science Foundation;
the Italian Istituto Nazionale di Fisica Nucleare; the Ministry of Education,
Culture, Sports, Science and Technology of Japan; 
the Natural Sciences and Engineering
Research Council of Canada; the National Science Council of the Republic of
China; the Swiss National Science Foundation; the A. P. Sloan Foundation; the
Bundesministerium fuer Bildung und Forschung, Germany; and the Korea Science
and Engineering Foundation.



\begin{thebibliography}{99}
\bibitem{pseudo} The pseudorapidity and transverse energy of a particle 
of energy $E$ and polar angle $\theta$ (measured from the proton 
beam direction) are defined as 
$\eta=-\ln(\tan\frac{\theta}{2})$ and $E_T=E\,\sin\theta$, respectively.
\end{thebibliography}
\end{document}